\documentclass{ws-ijmpcs}

\newcommand{\ke}{K_{e 2}}
\newcommand{\km}{K_{\mu 2}}

\newcommand{\kcpnn}{K^{+} \to \pi^{+} \nu \bar{\nu}}

\begin{document}

\markboth{V. Kozhuharov}
{RATIO OF THE CHARGED KAON LEPTONIC DECAYS AT NA62}

%
\catchline{}{}{}{}{}
%

\title{MEASUREMENT OF THE RATIO OF THE CHARGED KAON LEPTONIC DECAYS AT NA62}

\author{Venelin Kozhuharov\footnote{
Speaker, for
the NA62 Collaboration: F.~Ambrosino, A.~Antonelli, G.~Anzivino, R.~Arcidiacono, W.~Baldini, S.~Balev,
S.~Bifani, C.~Biino, A.~Bizzeti, B.~Bloch-Devaux
% ...
}}

\address{Laboratori Nazionali di Frascati - INFN, 40 E. Fermi, Frascati (Rome), Italy \\
Faculty of Physics, University of Sofia ``St. Kl. Ohridski'', 5 J. Bourchier Blvd., 
Sofia, Bulgaria \\
Venelin.Kozhuharov@cern.ch
}

\maketitle

\begin{history}
\received{30 11 2013}
\revised{30 11 2013}
\end{history}

\begin{abstract}
The ratio of the leptonic charged kaon decays $R_K= \Gamma(K^{\pm} \to e^{\pm}\nu) / \Gamma(K^{\pm} \to \mu^{\pm}\nu)$ is sensitive to the structure of the week interactions and can be precisely calculated within the Standard Model. 
Presence of New Physics can introduce a shift on its value of the order of a percent. 
The NA62 experiment at CERN SPS used data from a dedicated run in 2007 to perform a measurement of this ratio and probe the lepton universality. The data analysis technique and the final results are presented.

\end{abstract}

\ccode{PACS numbers:13.20.Eb, 11.30.Hv}

\section{Introduction}

Within the Standard Model the dilepton charged pseudoscalar meson decays proceed as tree level 
processes through a W exchange. 
However, the helicity conservation leads to a strong suppression of the electron mode.
The Standard Model (SM) expression for the ratio $R_K= \Gamma(Ke2) / \Gamma(K\mu 2)$ is a function 
of the masses of the participating particles and given by
\begin{equation}
R_K=\frac{m_e^2}{m_{\mu}^2} \left(  \frac{m_K^2 - m_e^2}{m_K^2 - m_{\mu}^2} \right)   (1+\delta R_K),
\end{equation}
where the term $ \delta R_K = -(3.79 \pm 0.04) \% $ represents the radiative corrections. 
In the ratio $R_K$ the theoretical uncertainties on the hadronic matrix element cancel 
resulting in an extremely precise prediction $R_K = (2.477 \pm 0.001) \times 10^{-5} $ \cite{ke2-thnew}. 

Due to the impossibility to distinguish the neutrino flavor the experimentally measured ratio is 
sensitive to possible lepton flavor violation effects. In particular, 
various LFV extensions of the SM (MSSM, different two Higgs doublet models) predict constructive or destructive contribution to $R_K$ as high as \%\cite{Masiero}

\section{The NA62 experiment at CERN SPS}

The NA62 experiment at CERN SPS is a continuation of the CERN long standing kaon physics program.
During its early stage, in 2007 and 2008, a modified setup of the NA48/2 experiment was used and the data 
taking was primarily devoted to the study of the $\ke$ decays.

The kaon beam was formed by a primary 400 GeV/c proton beam extracted from SPS hitting a 
400 mm long beryllium target. 
The secondary particles were selected with a momentum of $(74 \pm 1.4)$ GeV/c with the possibility 
to use simultaneous or single positive and negative beams. The kaon fraction in the beam was about 6\%. 

The decay products were registered by the NA48 detector\cite{bib:na48} which followed the
 114 m long evacuated decay region. 
The charged particles momentum was measured 
by a magnetic spectrometer housed in a helium tank at atmospheric pressure. 
It consisted of four drift chambers separated by a dipole magnet
 which provided a 265 MeV/c horizontal momentum kick, 
leading to a resolution $\sigma(p)/p = (0.48 \oplus 0.009p[GeV/c]) )\%  $ . 
Precise time information and trigger
condition was obtained from a scintillator hodoscope with time resolution of 150 ps which 
was followed by a 27 radiation lengths quasi-homogeneous liquid
krypton electromagnetic calorimeter (LKr), measuring photon and electron energy with a
resolution $\sigma(E)/E = 3.2\%/ \sqrt{E} \oplus 9\%/ E \oplus  0.42\%$, energy in GeV. It was also able to 
provide particle identification based on the energy deposit by different particles with respect to 
their momentum.


A lead bar with 9.2 radiation lengths was placed in front of the LKr during 55\% of the data taking 
to study the muon misidentification probability. Data with three different beam conditions were collected - 65\% 
only with $K^+$ beam, 8\% only with $K^-$, and the rest with simultaneous beams

 In 2009 the experimental apparatus was dismantled to allow the 
construction of the main stage of the experiment devoted to the 
study of the extremely rare  decay $\kcpnn$\cite{bib:na62tdr}.

\section{Event selection}

The final NA62 result\cite{bib:na62-ke2}.  supersedes 
the early NA62 result which was based on the analysis of a partial 
data sample collected in 2007\cite{bib:na62-ke2prel}. 

The ratio $R_K$ can be expressed as 
\begin{equation}
R_K =
\frac{1}{D}\cdot
\frac{N(\ke)-N_B(\ke)}
{N(\km) - N_B(\km)}
\cdot
\frac{A(\km)\times\epsilon_{\mathrm{trig}}(\km)\times f_\mu}
{A(\ke)\times\epsilon_{\mathrm{trig}}(\ke)\times f_e} \cdot \frac{1}{f_{LKr}},
\label{RKexp}
\end{equation}
where  $N(K_{\ell 2})$, $\ell=e,\mu$ is the number of the selected $\ke$ 
and $\km$ candidates,  
$N_B(K_{\ell 2})$ is the number of expected background events, 
$f_\ell$ is the efficiency for particle identification, 
$A(K_{\ell 2})$ is the geometrical efficiency for registration obtained from Monte Carlo simulation, 
$\epsilon_{\mathrm{trig}}$ is the trigger efficiency, $D=150$ is the downscaling factor for $\km$ events and $f_{lkr}$ is the global efficiency of the LKr readout. Both $f_\ell$ and $\epsilon_{\mathrm{trig}}$ are higher than 99\%. 

The analysis was performed in individual momentum bins for all the four different data samples - $K^+(noPb)$, $K^+(Pb)$, $K^-(noPb)$, $K^-(Pb)$ - resulting into 40 independent values for the $R_K$. 

The similarity between the two decays allowed to exploit systematics cancellations in the ratio by using common selection criteria. The events were required to have only one reconstructed charged track consistent with kaon decay within the detector geometrical acceptance with momentum in the interval $13 ~GeV/c < p < 65 ~GeV/c$. 
The background was additionally suppressed by vetoing events with clusters in the LKr not associated with the track with energy more than 2 $GeV$. 
The particle identification was based on the $E/p$ variable, where $E$ is the energy deposited in the LKr and $p$ is the momentum measured by the spectrometer. 
It had to be close to one for electrons and less than 0.85 for muons. 
Under the assumption of the particle type the missing mass squared was calculated $M_{miss}^2 = (P_K - P_l)^2$, 
where $P_K$ ($P_l$) is the kaon (lepton) four momentum. A momentum dependent cut on the $M_{miss}^2$  was used.

\section{Background estimation}

The dominant background contribution in the $\ke$ sample was identified to come from $\km$ events with muons leaving all their energy in the electromagnetic calorimeter. 
\begin{figure}[!htb]
    \resizebox{0.53\textwidth}{!}{\includegraphics[width=0.53\textwidth]{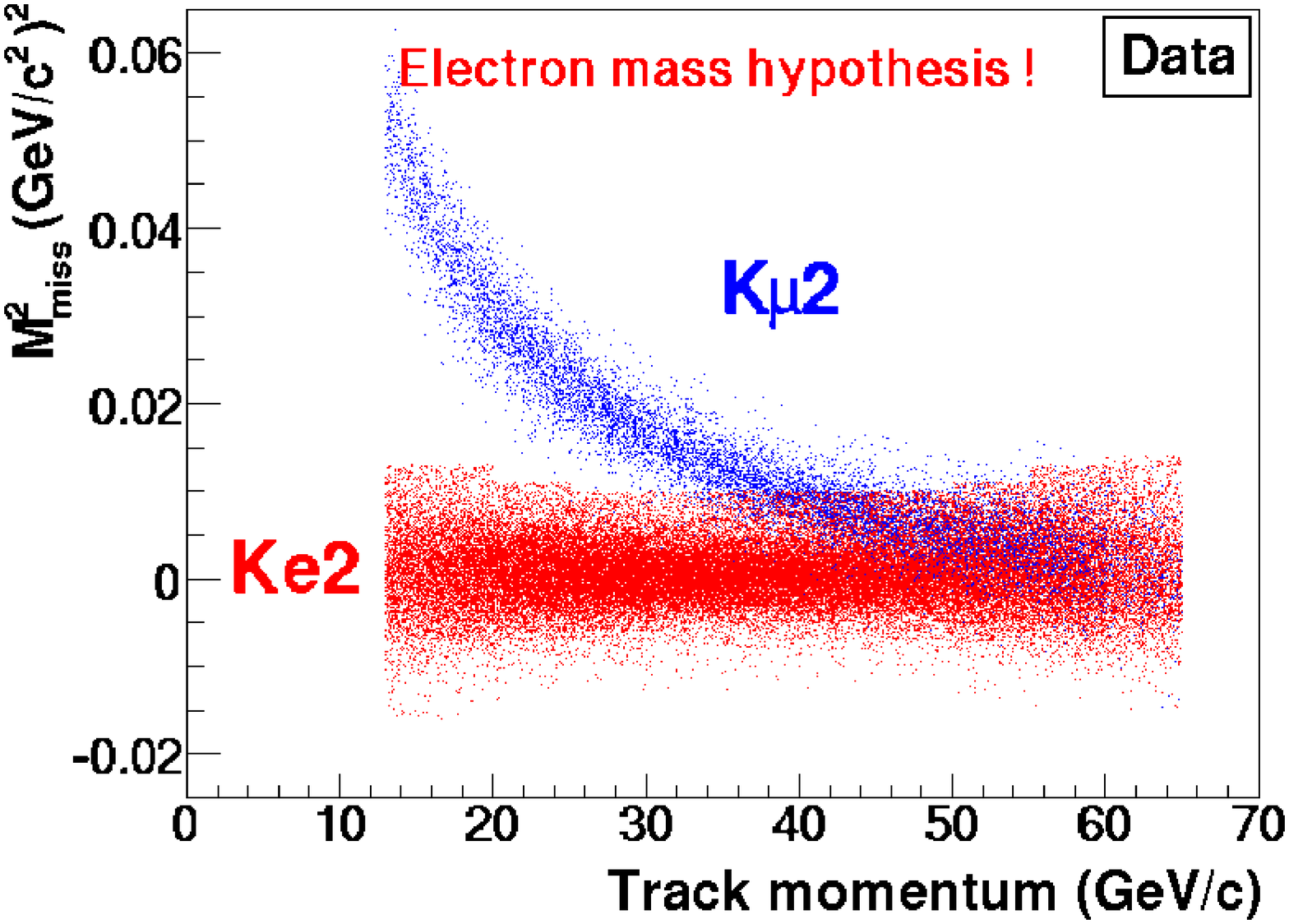}}
    \resizebox{0.4\textwidth}{!}{\includegraphics[width=0.4\textwidth]{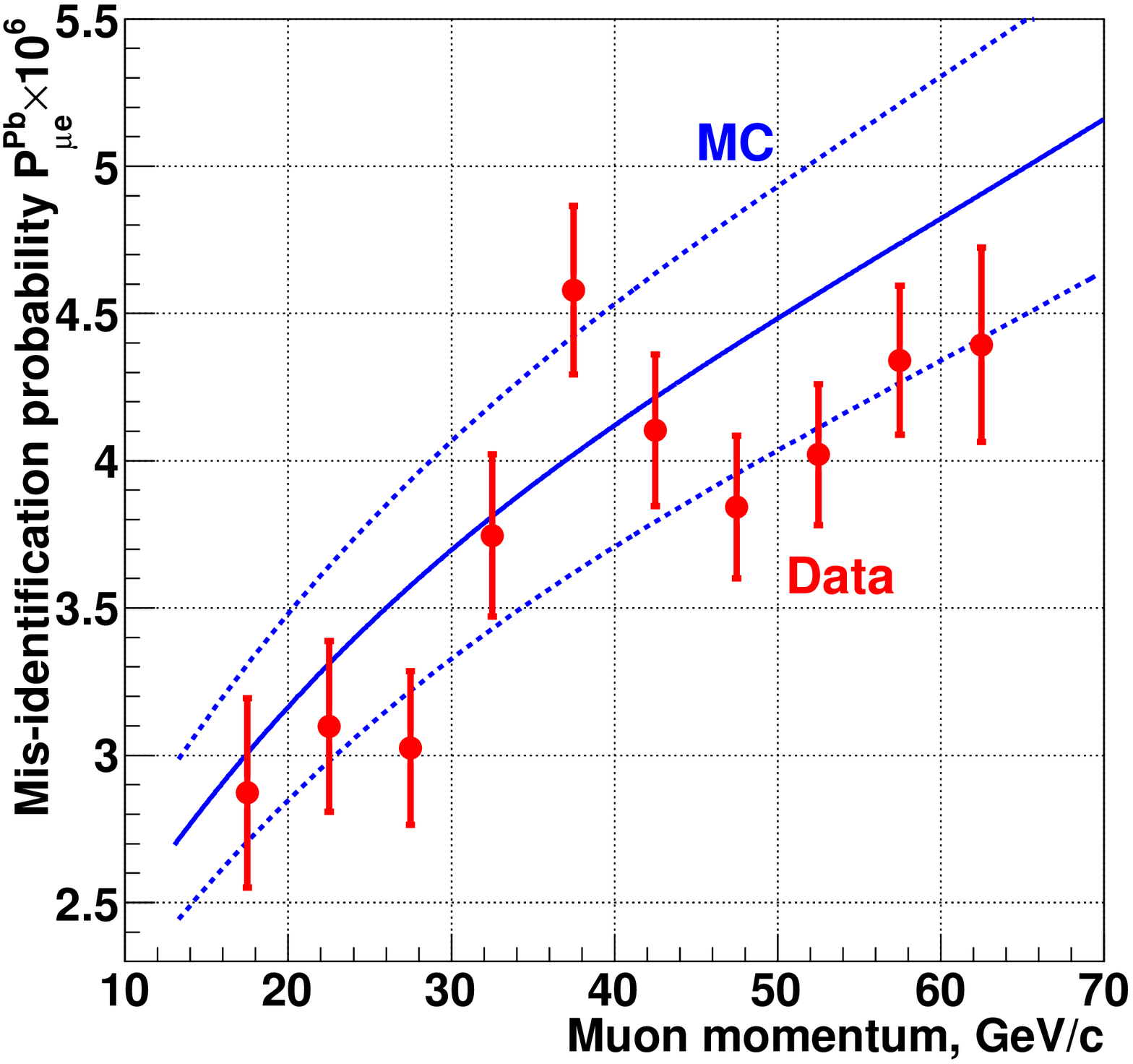}}
\put(-250,141){ \bf (a)}
\put(-80,141){ \bf (b)}
    \caption{ {(a) Missing mass distribution as a function of the charged track momentum for $\ke$ and $\km$ events under the electron hypothesis. (b) Probability to misidentify muon as an electron for muons traversing the lead bar compared to the Monte Carlo simulation. } \label{fig:ke2-back} }
\end{figure}
The two decays are well separated below 35 $GeV/c$ track momentum but completely overlap kinematically for higher values as seen in fig. \ref{fig:ke2-back}(a). 
A sample of data with a Pb bar placed in front of the LKr was used for the selection of clean muons. 
The study of the probility of a muon faking an electron was performed separately in the different momentum bins. 
The obtained misidentification probability $P_{\mu e}^{Pb}$ as a function of the muon momentum is shown in fig. \ref{fig:ke2-back}(b). It has to be corrected for the interactions (ionization energy loss and bremsstrahlung) 
of the muons inside the Pb bar. 
The correction factor was obtained by means of Geant4\cite{bib:geant4} simulation. 
The total background from $\km$ events was $(5.64 \pm 0.20)\%$ with uncertainty dominated by the statistics used for the determination of $P_{\mu e}^{Pb}$.

By definition $R_K$ includes only the inner bremsstrahlung 
component of the $K^{\pm} \to e^{\pm} \nu \gamma $ decays. 
The structure dependent and the interference terms were 
treated as background with the dominant being the $SD^+$ one since it mimics the $\ke$ signal once the 
photon is missed. The total contribution was $ (2.60 \pm 0.11)\%$ with uncertainty dominated by the experimental 
measurement of the decay rate\cite{bib:kloe-ke2g}.

At low track momentum the most significant background source was identified to be the muon halo 
with a contribution of $(2.11 \pm 0.09)\%$. 

The total background in the $\ke$ sample was $ (10.95 \pm 0.27)\%$. The missing mass distribution for the reconstructed $\ke$ data events together with the simulation of the signal and backgrounds are shown in fig. \ref{fig:ke2}(a). 

The background in the $\km$ sample was $(0.50 \pm 0.01)\%$ due to the beam halo. 

\section{Results and conclusions}

A total of $145958$ $\ke$ and $4.28\times 10^{7}$  $\km$ candidates were reconstructed. 
The major sources of the systematic uncertainty to $R_K$ are listed in table \ref{tab:rk-syst}. 
The dominant contribution is due to the background subtraction in the $\ke$ data sample. 
\begin{table}[!ht]
\tbl{Systematics contribution to $R_K$ ($\delta R_K*10^{5}$)}
{\begin{tabular}{@{}lc@{}} \toprule
$\km$ background in $\ke$ & 0.004 \\
$K^{\pm} \to e^{\pm} \nu \gamma $ ($SD^+$) background & 0.002\\
$K^{\pm} \to \pi^0 e^{\pm} \nu $ and $K^{\pm} \to \pi^+ \pi^0 $ backgrounds & 0.003 \\
Muon halo background & 0.002 \\
Spectrometer material composition & 0.002 \\
Acceptance correction & 0.002\\
Spectrometer alignment & 0.001\\
Electron identification inefficiency & 0.001\\
1-track trigger inefficiency & 0.001\\
LKr readout inefficiency & 0.001
\\ \colrule
Total systematics& 0.007\\
\botrule
\end{tabular} \label{tab:rk-syst}}
\end{table}

\begin{figure}[!htb]
    \resizebox{0.42\textwidth}{!}{\includegraphics[width=0.42\textwidth]{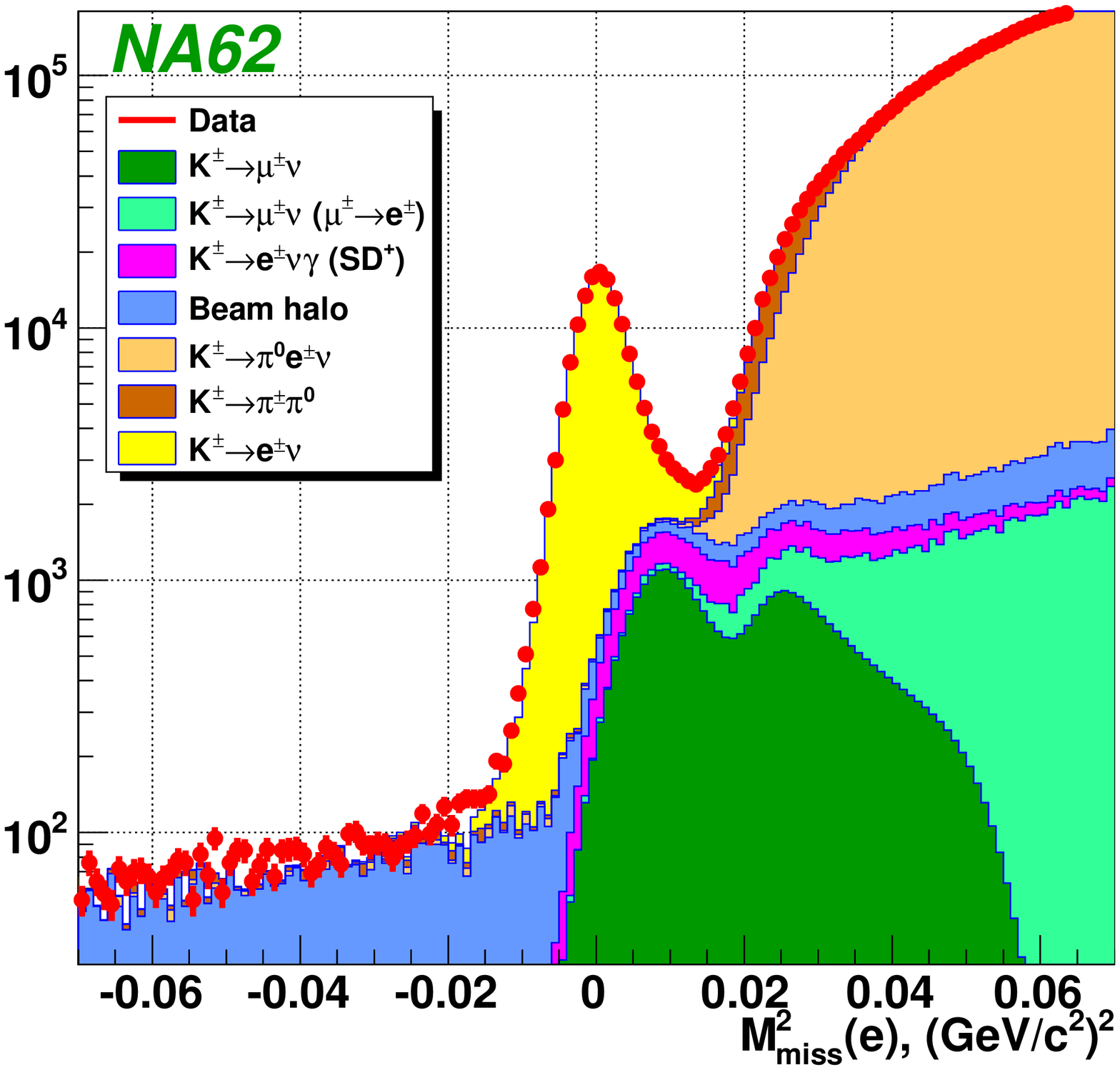}}
    \resizebox{0.50\textwidth}{!}{\includegraphics[width=0.50\textwidth]{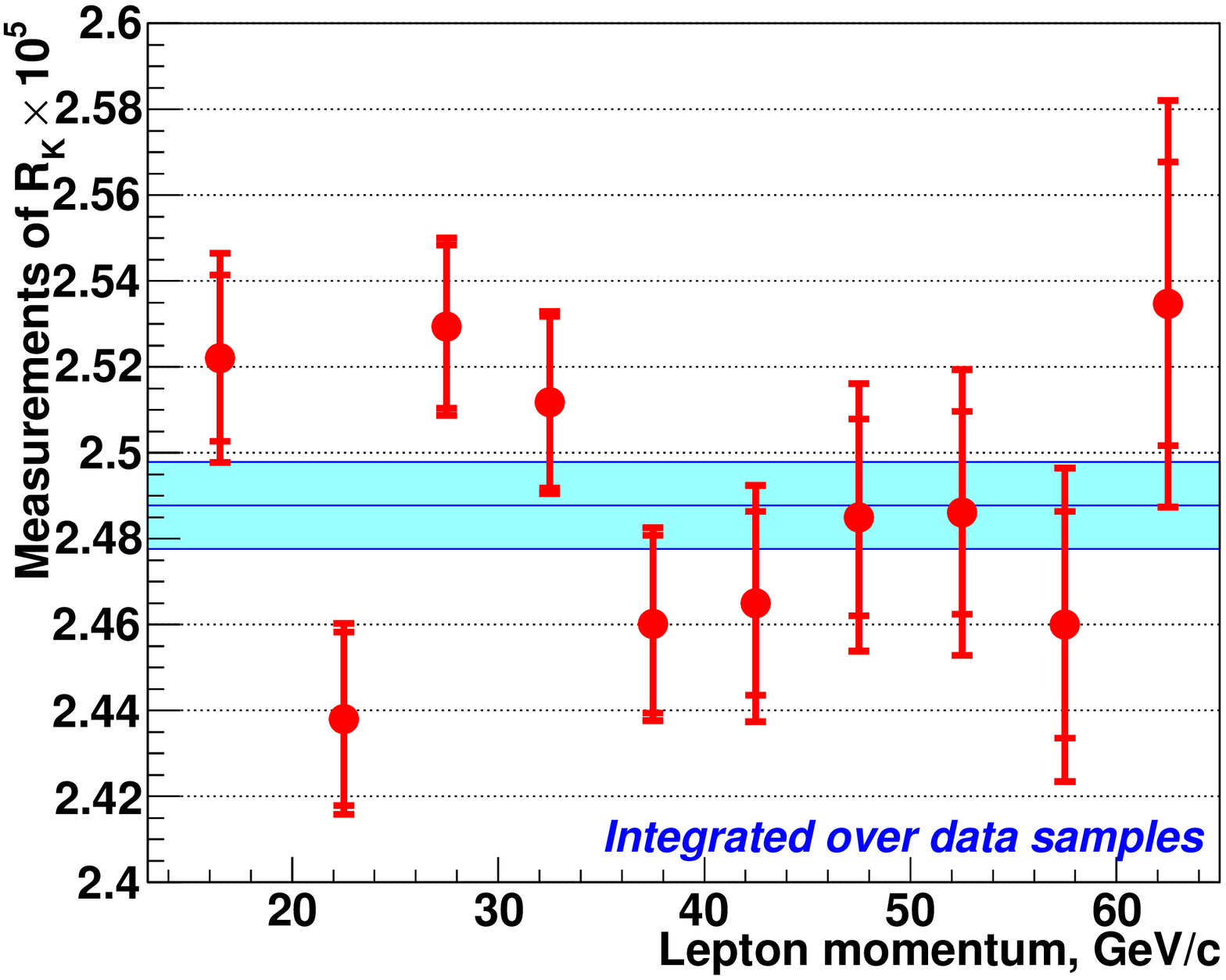}}
\put(-260,145){ \bf (a)}
\put(-100,145){ \bf (b)}
    \caption{ {(a) Squared missing mass distribution for the reconstructed data events (dots) together with the major background contribution. (b) $R_K$ in bins of track momentum integrated over data samples. The horizontal dashes represent the statistical only and the total error.} \label{fig:ke2} }
\end{figure}

The value of $R_K$ in the individual momentum bins integrated over 
the four different data samples is shown in fig. \ref{fig:ke2}(b). 
The final result
\begin{equation}
  R_K = (2.488 \pm 0.007_{stat} \pm 0.007_{syst})\times 10^{-5}.
\end{equation}
was obtained by a fit to the 40 independent $R_K$ values with $\chi^2/$ndf = 47/39
and is the most precise measurement to date. 
It is consistent with the Standard Model prediction and with the present PDG value\cite{bib:ke2-pdg}. 

The main data taking of the NA62 experiment will start in 2014. 
The huge kaon flux ($\sim 10^{13}$ kaon decays in the fiducial volume) together with the excellent resolution, particle identification, and veto capabilities could allow to push down the uncertainty in the $R_K$ measurement to 0.2\%. 
The simultaneous kaon and pion beams, provided that precise knowledge of the beam composition is achievable, could also be used to perform the first measurement of ratio  $R^e_{K\pi} = \Gamma(Ke2)/\Gamma(\pi e2)$.

\end{document}